\documentclass[apj,preprint]{emulateapj}
\usepackage{mathptmx}
\usepackage{subfigure}  % use for side-by-side figures
\usepackage{hyperref}
\hypersetup{citecolor=green, linkcolor=red}
\usepackage{amsmath}
\usepackage[normalem]{ulem}
\def\gtorder{\mathrel{\raise.3ex\hbox{$>$}\mkern-14mu
             \lower0.6ex\hbox{$\sim$}}}
\def\ltorder{\mathrel{\raise.3ex\hbox{$<$}\mkern-14mu
             \lower0.6ex\hbox{$\sim$}}}

\slugcomment{Accepted by \apj\ on August 18, 2015}

\shorttitle{Precursors prior to SNe IIb}

\shortauthors{Strotjohann et al.}

\begin{document}

\title{Search for precursor eruptions among Type IIb supernovae}
%\title{Absence of precursor eruptions prior to Type IIb supernovae}
\author{Nora L. Strotjohann\altaffilmark{1,2},
Eran O. Ofek\altaffilmark{1},
Avishay Gal-Yam\altaffilmark{1},
Mark Sullivan\altaffilmark{3},
Shrinivas R. Kulkarni\altaffilmark{4},
Nir J. Shaviv\altaffilmark{5,6},
Christoffer Fremling\altaffilmark{7},
Mansi M. Kasliwal\altaffilmark{8},
Peter E. Nugent\altaffilmark{9,10},
Yi Cao\altaffilmark{4},
Iair Arcavi\altaffilmark{11,12},
Jesper Sollerman\altaffilmark{7},
Alexei V. Filippenko\altaffilmark{10},
Ofer Yaron\altaffilmark{1},
Russ Laher\altaffilmark{13}, and
Jason Surace\altaffilmark{13}
}
\altaffiltext{1}{Benoziyo Center for Astrophysics, Weizmann Institute of Science, 76100 Rehovot, Israel}
\altaffiltext{2}{Desy Zeuthen, 15738 Zeuthen, Germany}
\altaffiltext{3}{School of Physics and Astronomy, University of Southampton, Southampton SO17 1BJ, UK}
\altaffiltext{4}{Cahill Center for Astronomy and Astrophysics, California Institute of Technology, Pasadena, CA 91125, USA}
\altaffiltext{5}{School of Natural Sciences, Institute for Advanced Study, 1 Einstein Dr, Princeton, NJ 08540, USA}
\altaffiltext{6}{Racah Institute of Physics, Hebrew University, Jerusalem 91904, Israel}
\altaffiltext{7}{The Oskar Klein Centre, Department of Astronomy, Stockholm University, AlbaNova, 10691 Stockholm, Sweden}
\altaffiltext{8}{Observatories of the Carnegie Institution for Science, 813 Santa Barbara St., Pasadena, CA 91101, USA}
\altaffiltext{9}{Lawrence Berkeley National Laboratory, 1 Cyclotron Road, Berkeley, CA 94720, USA}
\altaffiltext{10}{Department of Astronomy, University of California, Berkeley, CA 94720-3411, USA}
\altaffiltext{11}{Las Cumbres Observatory Global Telescope Network, 6740 Cortona Dr., Suite 102, Goleta, CA 93111, USA}
\altaffiltext{12}{Kavli Institute for Theoretical Physics, University of California, Santa Barbara, CA 93106, USA}
\altaffiltext{13}{Spitzer Science Center, California Institute of Technology, M/S 314-6, Pasadena, CA 91125, USA}

\begin{abstract}
The progenitor stars of several Type IIb supernovae (SNe) show indications for extended hydrogen envelopes. These envelopes might be the outcome of luminous energetic pre-explosion events, so-called precursor eruptions.
We use the Palomar Transient Factory (PTF) pre-explosion observations of a sample of 27 nearby Type IIb SNe to look for such precursors during the final years prior to the SN explosion. No precursors are found when combining the observations in 15-day bins, and we calculate the absolute-magnitude-dependent upper limit on the precursor rate.
At the 90\% confidence level, Type IIb SNe have on average $<0.86$ precursors as bright as absolute $R$-band magnitude $-14$ in the final 3.5 years before the explosion and $<0.56$ events over the final year. In contrast, precursors among SNe IIn have a $\gtrsim 5$ times higher rate. 
The kinetic energy required to unbind a low-mass stellar envelope is comparable to the radiated energy of a few-weeks-long precursor which would be detectable for the closest SNe in our sample. Therefore, mass ejections, if they are common in such SNe, are radiatively inefficient or have durations longer than months. Indeed, when using 60-day bins a faint precursor candidate is detected prior to SN\,2012cs ($\sim2$\% false-alarm probability).
We also report the detection of the progenitor of SN\,2011dh which does not show detectable variability over the final two years before the explosion. The suggested progenitor of SN\,2012P is still present, and hence is likely a compact star cluster, or an unrelated object.

\end{abstract}


\section{Introduction}
\label{sec:introduction}

Violent stellar activity, as for example precursor eruptions, during the years prior to the terminal explosion of a star as a supernova (SN) can give insights into the properties of progenitor stars at this crucial stage of their life. In case these eruptions are causally connected to the final explosion they might even provide hints about the explosion mechanism. Indications for such luminous flares were reported previously, mainly among Type IIn SNe (e.g., \citealt{foley2007, pastorello2007, fraser2013, ofek2013, ofek2014b, mauerhan2013}), and a possible precursor was also reported for a Type Ic SN (\citealt{corsi2014}). Here we search for precursor eruptions, i.e., small explosions prior to the SN, in the pre-explosion light curves of Type IIb SN.

The defining property of SNe~IIb, the first known example of which was SN\,1987K (\citealt{filippenko1988}), is their spectral evolution. The spectrum first resembles that of a hydrogen-rich Type II SN and later turns into a helium-dominated spectrum similar to that of a Type Ib SN (e.g., \citealt{filippenko1993, filippenko1994, chornock2011, ben-ami2015}; see \citealt{filippenko1997} for a general review of SN spectra). This evidence suggests that the progenitors of SNe IIb likely consist of a helium core surrounded by a low-mass hydrogen envelope that expands and consequently the hydrogen features fade away within a few weeks (e.g., \citealt{woosley1994, tsvetkov2009, chevalier2010, bersten2012, ergon2014a, ergon2014b, marion2014, nakar2014}). 

Another characteristic is that some SNe~IIb show two optical photometric peaks in their light curve (e.g., SN\,1993J, \citealt{wheeler1993}; SN\,2011dh, \citealt{arcavi2011}; SN\,2013df, \citealt{vandyk2014}).  The first peak can be attributed to thermal emission from a shock-heated extended hydrogen envelope (\citealt{bersten2012, nakar2014}), while the second one is presumably powered by radioactivity.
\cite{nakar2014} suggest that the envelope's radius and mass can be inferred from the shape of the first peak. They have also shown that standard red giant or Wolf-Rayet progenitors cannot explain the observed light curves. Indeed, large photospheric radii of several hundred solar radii have been measured for several SNe~IIb at early times (e.g., \citealt{woosley1994, tsvetkov2009, bersten2012, horesh2013, ergon2014a, ergon2014b}).

Additionally, an ultraviolet (UV) excess relative to the spectral shape in the optical has been observed for several SNe~IIb whose UV spectra were obtained within the first month after explosion. It can be explained by the presence of a dense circumstellar medium (CSM) above the photosphere (\citealt{ben-ami2015}). Double-peaked light curves and excess UV radiation are, however, not observed for all SNe~IIb (\citealt{arcavi2015, ben-ami2015}). Moreover, \cite{chevalier2010} argue that some SNe~IIb show indications for much more compact progenitors having $R\approx 1\,R_\sun$.

It is unclear how extended envelopes or dense CSM shells are created. They could consist either of material in hydrodynamic equilibrium (\citealt{benvenuto2013,bersten2012}) or of unbound material ejected during the interaction with a binary companion (e.g., \citealt{chevalier2012}), by a stellar wind, or by an instability related to the final stages of stellar burning (e.g., \citealt{rakavy1967, arnett2011a, arnett2011b, quataert2012, shiode2014}). Such an instability can generate a precursor event that can release some of its energy in visible light.

A systematic search for precursor eruptions among SNe~IIn has previously been performed by \cite{ofek2014b}, with six likely precursor explosions observed in a sample of 16 objects. A quantitative analysis showed that, assuming a uniform population, at the 99\% confidence level at least 98\% of the SNe~IIn exhibit one or several precursor events brighter than an absolute magnitude of $-14$ within the last 2.5 years prior to their explosion. These results apply to all SNe~IIn under the assumption that they form a homogeneous population which is well represented by the used sample. Furthermore, \cite{ofek2014b} find possible correlations between the amount of energy radiated in a precursor event and the SN rise time, peak magnitude, and radiated energy. If real, the correlation is consistent with the idea that SNe IIn are powered mainly by the interaction of the SN ejecta with a massive CSM created in precursor eruptions.

Only SNe having strong and persistent narrow hydrogen lines were included in the sample of \cite{ofek2014b}. It was found recently that some core-collapse SNe, dubbed "flash-spectroscopy SNe", exhibit narrow hydrogen lines within the first days after the explosion which vanish subsequently (\citealt{gal-yam2014,khazov2015}). This evolution might indicate that a dense CSM shell or an envelope is located only in the immediate vicinity of the progenitor star and is swept up by the ejecta shortly after the explosion. Owing to the short duration of the CSM interaction, these objects are not considered in the sample of SNe~IIn of \cite{ofek2014b} and might not undergo precursor explosions as frequently. We note that the detection of flash-spectroscopy signatures provide further evidence for mass ejections prior to SN explosions.

In another search for precursors among SNe~IIn by \cite{bilinski2015}, no pre-explosion events were found. However, with six objects their sample is smaller, and they do not provide an absolute-magnitude-dependent upper limit on the precursor rates, so a close comparison is not possible at this stage.

Here, we extend the search of \cite{ofek2014b} to Type IIb SNe. In \S\ref{sec:sample} we explain how the SN sample was selected, and in \S\ref{sec:observations} we describe the observations used in this analysis. Our search for precursor explosions is presented in \S\ref{sec:search}. The pre-explosion observations of three nearby SNe are evaluated in more detail in \S\ref{sec:individual}. In \S\ref{sec:rates}, we calculate the sample control time and derive an upper limit on the precursor rate of SNe~IIb. Section \ref{sec:discussion} discusses whether the ejection of a low-mass stellar envelope is likely detectable in this search, and \S\ref{sec:summary} summarizes the results.\\

\section{Sample Selection}
\label{sec:sample}

The sample is selected by initially considering all nearby SNe IIb detected either by the Palomar Transient Factory (PTF; \citealt{law2009, rau2009}) or announced in Astronomer's Telegrams or IAU circulars since 2009 (i.e., after the start of the PTF project). SNe IIb are chosen based on their spectral evolution, and the main criterion is the appearance of helium features at $5876$ \AA, $6678$ \AA, and $7065$ \AA\ a few weeks after the explosion. Tools such as the Supernova Identification Code (SNID; \citealt{blondin2007}) and Superfit (\citealt{howell2005}) are used to compare our spectra to the spectra of known SNe IIb in case the spectral features are not obvious. We note that in general, our SN classification is based on human decision and hence may be biased. To increase the chances of detecting faint precursors, we restrict ourselves to nearby SNe with $z\leqslant0.05$, corresponding to a luminosity distance of $220$ Mpc.
Only SNe having a large number of PTF observations prior to their explosion are selected. We require about 20 images (either before or long after the SN explosion) to construct a high-quality reference image, and another 20 science images before the explosion date.

\begin{deluxetable*}{llllllllllll}
\tablecolumns{11}
\tablewidth{\textwidth}
\tablecaption{Supernova Sample}
\tablehead{
\colhead{Name}          &
\colhead{$\alpha$(J2000)}      &
\colhead{$\delta$(J2000)}     &
\colhead{$z$}             &
\colhead{DM}            &
\colhead{$E(B-V)$}	&
\colhead{$M_{R,{\rm peak}}$}  &
\colhead{$t_0$} &
\colhead{$t_{{\rm peak}}$} &
\colhead{FAP} &
\colhead{DP} &
\colhead{ref. period} \\
\colhead{}    &
\colhead{(deg)} &
\colhead{(deg)} &
\colhead{}    &
\colhead{(mag)} &
\colhead{(mag)} &
\colhead{(mag)} &
\colhead{(day)} &
\colhead{(day)} &
\colhead{} &
\colhead{} &
\colhead{}
}
\startdata
PTF\,09dxv 		& 347.144705	& $+$18.937131	& 0.0322	& 35.77 & 0.145		 & $-$18.0$^*$   & 55079	& 55094  & 0.000 & & $>$56041 \\
SN\,2009nf, PTF\,09gyp 	& 029.736494	& $-$07.282473  & 0.046		& 36.57 & 0.027		 & $-$17.6 	& 55138		& 55151  & 0.025 & & $>$55835 \\
PTF\,09hnq 		& 345.470095	& $+$14.413534  & 0.027		& 35.38 & 0.108		 & $-$17.5$^*$	& 55157		& 55168  & 0.002 & & $>$56449 \\
PTF\,09ism 		& 176.149461	& $+$10.212143	& 0.03		& 35.61 & 0.075 	 & $-$17.4	& 55194	& 55200  & 0.000 & & $>$56328 \\
PTF\,10fqg 		& 190.457745	& $+$11.591142 & 0.0278		& 35.44 & 0.032 	 & $-$16.6 	& 55302	& 55323  & 0.126 & & $>$56712 \\
PTF\,10qrl 		& 347.470125	& $+$13.132566  & 0.0396	& 36.23 & 0.076 	 & $-$17.0 	& 55413	& 55427  & 0.027 & & $>$56503 \\
PTF\,10tzh 		& 257.305622	& $+$41.755139	& 0.034		& 35.89 & 0.025 	 & $-$16.3$^*$   & 55437	& 55459  & 0.002 & y & $>$56360 \\
PTF\,10xfl 		& 034.955874	& $+$15.295001	& 0.05		& 36.76 & 0.105 	 & $-$18.2$^*$ & 55456	& 55470  & 0.030 & & $>$56533 \\
%PTF\,10aako 		& 		& $+$		& $-$	   & 0.0	& 	 & 55		& 55	  & \\
SN\,2011dh, PTF\,11eon 	& 202.521152	& $+$47.169782 & 0.001683 	& 29.45 & 0.035 	 & $-$17.1  & 55713 	& 55732  & 0.050 & y & $>$56662 \\ %0.31 \\
SN\,2011hg, PTF\,11pdj	& 347.953215	& $+$31.016708	& 0.0236	& 35.08 & 0.075 	 &    		& 55856	&   & 0.000 & & $>$56735 \\
PTF\,11qju 		& 213.862526	& $+$36.408608	& 0.0282	& 35.48 & 0.089 	 & $-$17.5   & 55880	& 55902  & 0.007 & & $>$56324 \\
SN\,2012P, PTF\,12os	& 224.996177	& $+$01.890051	& 0.004533	& 32.04 & 0.051 	 & $-$16.3 & 55931	& 55952  & 0.006 & & $>$56710\\
PTF\,12fxj 		& 027.004256	& $+$35.708841	& 0.0148	& 34.06 & 0.056 	 & $-$17.4 & 55089	& 55102  & 0.054 & & $<$55127 \\
SN\,2012ey, PTF\,12iqw 	& 036.210492	& $+$16.181395	& 0.027		& 35.38 & 0.179 	 & $-$15.8$^*$   & 56183	& 56198  & 0.000 & & $<$55503 \\
PTF\,12jaa 		& 338.364087	& $+$00.740223	& 0.0237	& 35.09 & 0.087 	 & $-$15.6$^*$   & 56190	& 56193  & 0.000 & y & $<$55896\\
PTF\,13nu 		& 183.267741	& $+$32.613898	& 0.026		& 35.30 & 0.013 	 & $-$17.7   & 56356	& 56376  & 0.355 & & $<$55896 \\
SN\,2013bb, PTF\,13aby	& 213.058377	& $+$15.842080	& 0.01755	& 34.43 & 0.015 	 & $-$16.5   & 56369	& 56394  & 0.217 & & $>$56766 \\ %0.10 & \\
PTF\,13ajn 		& 129.528692	& $+$66.526227	& 0.03		& 35.62 & 0.049 	 & $-$17.3$^*$   & 56385	& 56404  & 0.077 & y & $<$55183 \\ %\\6 & y \\
& 	& 	& 	& 	 & 	 & 	& 	& 	& 	 & 	 & \& $<$56325 \\ %\\6 & y \\

SN\,2013cu, PTF\,13ast 		& 218.495242	& $+$40.239672	& 0.0258	& 35.28 & 0.012 	 & $-$18.5   & 56414	& 56426  & 0.003  & & $<$55042 \\ %\\6 & y \\
PTF\,13ebs 		& 140.284454	& $+$49.592651	& 0.027		& 35.38 & 0.022 	 & $-$16.9$^*$   & 56606	& 56622  & 0.003 & & $>$56780 \\
\hline\\
[-4ex]\\
SN\,2011ef 		& 352.737583 & $+$15.490083 & 0.0134		& 33.84 & 0.067 	 & 	  & 55760	&    & 0.095 & & $>$56406 \\
SN\,2012an 		& 261.042625	& $+$59.001916	& 0.0111	& 33.43 & 0.030 	 & 	  & 55978	&    & 0.000 & & $>$56778 \\
SN\,2012cs 		& 232.990208 & $+$68.245222 & 0.0218		& 33.56 & 0.027 	 & 	  & 56053	&    & 0.041 & & $<$55052 \\
LSQ\,12fwb 		& 006.355083	& $+$06.707194	& 0.03		& 35.62 & 0.024 	 & 	  & 56233	&    & 0.114 & & $<$55430 \\
LSQ\,12htu 		& 152.904625	& $-$07.386555	& 0.04		& 36.26 & 0.034 	 & 	  & 56282	&    & 0.006 & & $>$56712 \\
SN\,2013df 		& 186.622208	& $+$31.227305	& 0.002388	& 31.23 & 0.019 	 & $-$16.8	  & 56447	& 56470   & 0.965 & y & $<$56388 \\ %DM=31.10 in 1312.3984
PS1-14od 		& 050.275958 & $-$07.282611 & 0.02		& 34.72	 & 0.059	 & 	  & 56713	&    & 0.000 & & $<$55089
\enddata
\tablecomments{The SN sample. $\alpha$(J2000) and $\delta$(J2000) are the J2000.0 right ascension and declination, respectively; $z$ is the SN redshift obtained from spectroscopy. DM, the distance modulus, is derived from the redshift with $H_0=69.33$\,km\,s$^{-1}$\,Mpc$^{-1}$, $\Omega_M=0.24$, and $\Omega_\Lambda$=0.71 (\citealt{hinshaw2013}). The only exceptions are the three closest SNe and SN\,2012cs, where redshift-independent distance measurements of the host galaxies are available on NED. $E(B-V)$ is the Galactic extinction taken from \cite{schlegel1998}. $M_{R, {\rm peak}}$ is the absolute $R$-band magnitude of the brightest detection; asterisks indicate that the peak is not well observed and the SN might be considerably brighter.
$t_0$ is the MJD of the approximate explosion date estimated by picking a date between the last nondetection and the first detection; thus, for some SNe the uncertainty in $t_0$ can be many days. $t_{{\rm peak}}$ is the MJD of the brightest detection, where only the second peak is considered in case of a double-peaked light curve. FAP, the false-alarm probability, is the probability of detecting a false precursor candidate by coadding images in 15-day bins as estimated using the bootstrap method (see \S\ref{sec:methods}). The penultimate column (DP) indicates whether two peaks are observed in the light curve, and the last column specifies the chosen reference period. Below the solid line we list SNe added to our sample from the literature.
\\
{\it References:}\\
PTF\,09hnq, PTF\,10fqg, PTF\,10qrl, PTF\,10tzh, PTF\,10xfl, PTF\,11qju, PTF\,12jaa, PTF\,13nu, PTF\,13ajn, and PTF\,13ebs: reported here for the first time.\\
PTF\,09dxv: \cite{arcavi2010}.\\
SN\,2009nf (PTF\,09gyp): \cite{drake2009b, arcavi2010}.\\
PTF\,09ism: \cite{arcavi2010}. \\
SN\,2011dh (PTF\,11eon): \cite{arcavi2011, griga2011}. \\ %0.31 \\
SN\,2011hg (PTF\,11pdj): \cite{ciabattari2011, gal-yam2011, marion2011, tomasella2011}.\\
SN\,2012P (PTF\,12os): \cite{arcavi2012, borsato2012, dimai2012}.\\
SN\,2013bb (PTF\,13aby): \cite{howerton2013, elias-rosa2013}.\\ %0.10 & \\
SN\,2011ef: \cite{blanchard2011, parrent2011}.\\
SN\,2012an: \cite{chen2012, jha2012, newton2012}. \\
SN\,2012cs: \cite{rich2012}. \\ %\cite{marion2012, lipunov2012}. \\
SN\,2012ey: \cite{howerton2012, turatto2012}.\\
LSQ\,12fwb: \cite{hadjiyska2012}. \\
LSQ\,12htu: \cite{leguillou2012}. \\
SN\,2013cu: \cite{gal-yam2014}.\\
SN\,2013df: \cite{ciabattari2013, vandyk2013, vandyk2014, morales-garoffolo2014, ben-ami2015}.\\
PS1-14od: \cite{campbell2014}.\\
}
\label{tab:Samp}
\end{deluxetable*}

\begin{deluxetable*}{llllllll}[tb]
\tablecolumns{8}
\tablewidth{\textwidth}
\tablecaption{PTF Observations\label{tab:obs}}
\tablehead{
\colhead{Name}          &
\colhead{MJD--$t_0$}          &
\colhead{MJD}      &
\colhead{$m_{{\rm PTF}, R}$}     &
\colhead{$m_{{\rm PTF}, R}$ Err}  &
\colhead{Lim Mag}             &
\colhead{Flux}            &
\colhead{Flux Err} \\
\colhead{}    &
\colhead{(day)}    &
\colhead{(day)} &
\colhead{(mag)} &
\colhead{(mag)} &
\colhead{(mag)}    &
\colhead{(counts)} &
\colhead{(counts)}
}
\startdata
PTF\,09dxv & $-47.876$ & 55031.304 &  25.34 &  31.38   & 21.09       &  2.7  &     76.9\\
PTF\,09dxv & $-47.805$ & 55031.375 &  22.72 &   1.24   & 21.39       & 51.1  &     58.4\\
PTF\,09dxv & $-45.711$ & 55033.469 &  78.65 &  -       & 21.38       & $-9.0$  &     59.2\\
PTF\,09dxv & $-45.705$ & 55033.475 &  21.69 &   0.49   & 21.37       & 133.2 &     59.7\\
PTF\,09dxv & $-43.762$ & 55035.418 &  79.45 &  -       & 21.46       & $-18.8$ &     54.9\\
... 
\enddata
\tablecomments{Flux residuals in the pre-explosion light curve of all SNe in our sample. Magnitudes are calculated as ``asinh magnitudes'' (\citealt{lupton1999}), and have a meaning only when smaller than the limiting magnitude. The limiting magnitude here is at the $3\sigma$ level. This table is published in its entirety in the electronic version of ApJ. A portion of the full table is shown here for guidance regarding its form and content.
}
\label{tab:obs}
\end{deluxetable*}

\begin{figure*}[p]
\centerline{\includegraphics[width=18cm]{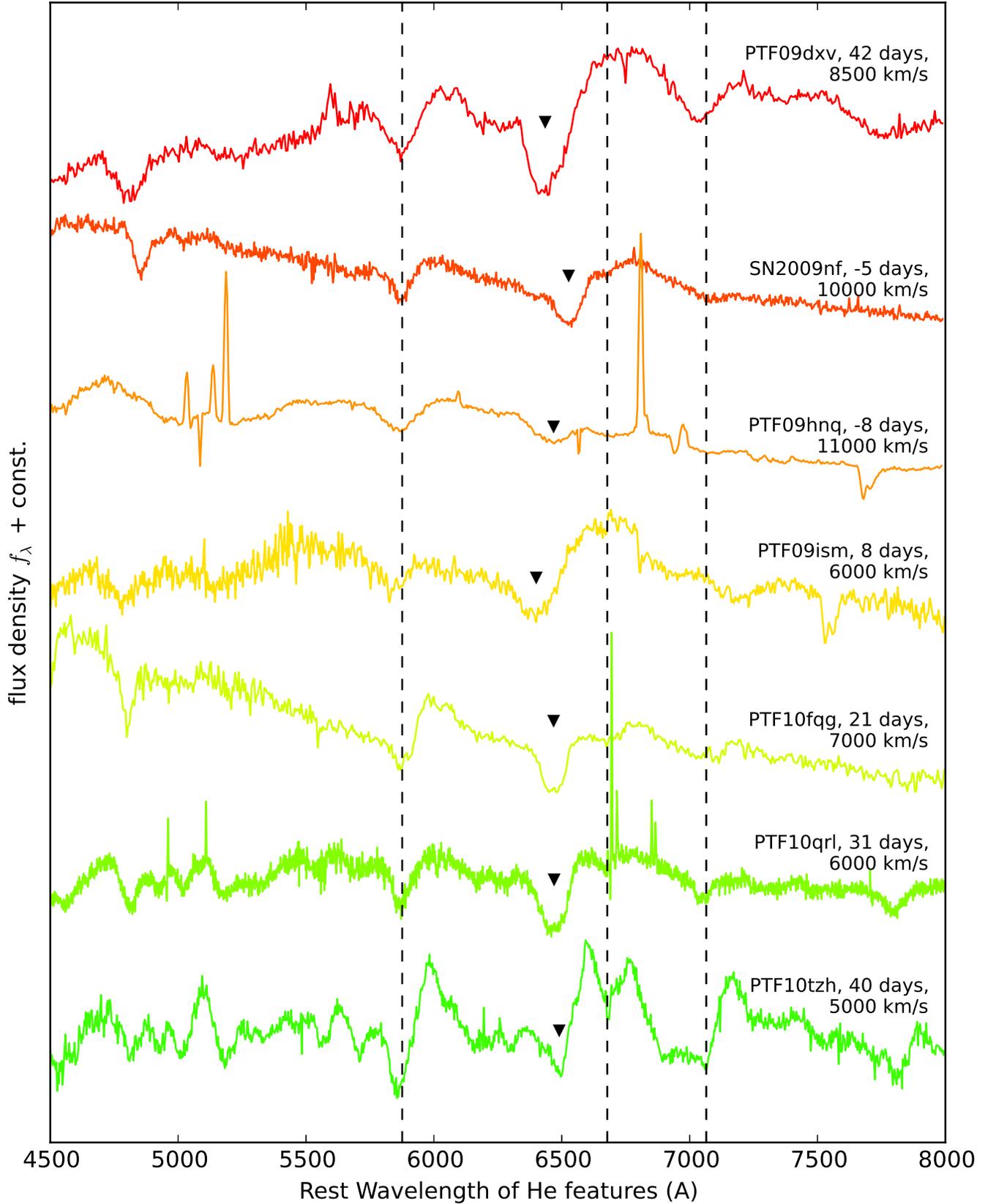}}
\caption{Representative spectra of the PTF-detected SNe in our sample. All spectra are redshifted by the indicated velocity relative to the rest frame such that their characteristic helium absorption features are at their rest wavelengths at $5876$ \AA, $6678$ \AA, and $7065$ \AA\ (indicated by broken lines). In addition, the black triangles mark the prominent H$\alpha$ absorption lines. The H$\alpha$ emission line is found to the right of these features and often has a flat-top profile owing to the helium absorption at $6678$ \AA. Some SNe are compared to known SNe~IIb identified with SNID (\citealt{blondin2007}) and the time after peak is indicated next to all spectra. All spectra are available electronically via the WISeREP webpage (\citealt{yaron2012}). Further spectra are shown in Figures \ref{fig:spectra2} and \ref{fig:spectra3}. \label{fig:spectra1}}
\end{figure*}

\begin{figure*}[p]
\centerline{\includegraphics[width=18cm]{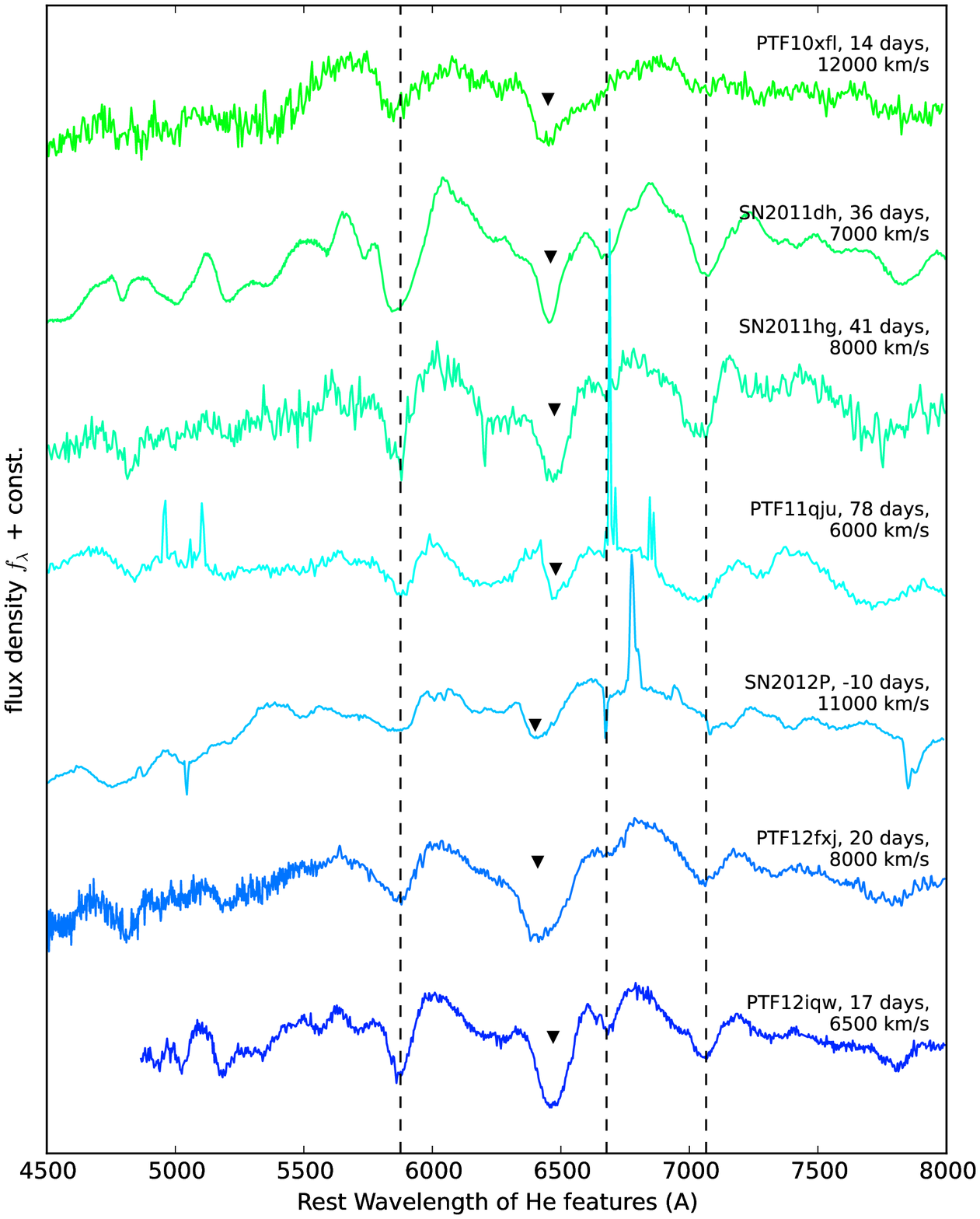}}
\caption{Figure \ref{fig:spectra1}, continued. \label{fig:spectra2}}
\end{figure*}

\begin{figure*}[p]
\centerline{\includegraphics[width=18cm]{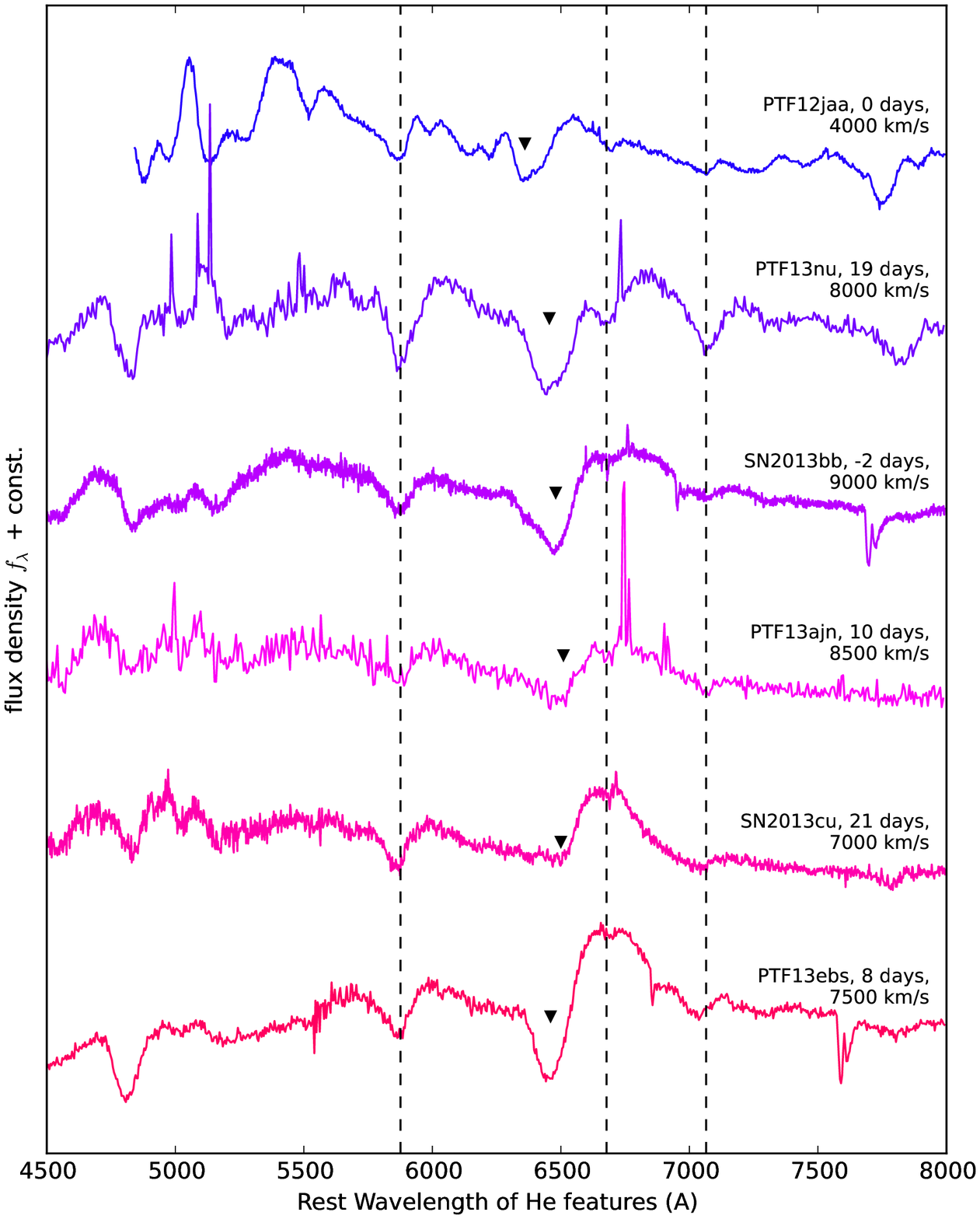}}
\caption{Figure \ref{fig:spectra1} and \ref{fig:spectra2}, continued. \label{fig:spectra3}}
\end{figure*}

Our final sample consists of 27 SNe listed in Table \ref{tab:Samp}. The majority were discovered by PTF, and additional SNe were found by the Lick Observatory Supernova Search (LOSS; \citealt{li2000, filippenko2001}), the Puckett Observatory Supernova Search\footnote{http://www.cometwatch.com/supernovasearch.html}, the La Silla Quest survey (LSQ; \citealt{baltay2013}), the Catalina Real-Time Transient Survey (CRTS; \citealt{drake2009a}), the Italian Supernova Search Project\footnote{http://italiansupernovae.org/} (ISSP), and the Panoramic Survey Telescope and Rapid Response System (Pan-STARRS; \citealt{hodapp2004}).
For three SNe (SN\,2011ef, SN\,2012an, and LSQ\,12fwb), no spectra are publicly available and we rely on the classification published in Astronomer's Telegrams and Central Bureau Electronic Telegrams by \cite{blanchard2011}, \cite{chen2012}, and \cite{hadjiyska2012}. Representative spectra of all SNe discovered by PTF are shown in Figures \ref{fig:spectra1}-\ref{fig:spectra3} and are electronically available from the WISeREP webpage\footnote{http://www.weizmann.ac.il/astrophysics/wiserep/} (\citealt{yaron2012}). The spectra were acquired using various facilities which are listed in Table \ref{tab:spec}.

Our sample includes SN\,2013cu (PTF\,13ast), which has been classified as a flash-spectroscopy SN (\citealt{gal-yam2014, khazov2015}). After the initial CSM interaction, SN\,2013cu evolves into a SN~IIb (see the spectrum in Figure \ref{fig:spectra3}) and is thus included in this sample. Out of all core-collapse SNe having spectra within the first ten days after their last nondetection, $\sim14$\% show flash-spectroscopy signatures (\citealt{khazov2015}). It is therefore likely that other SNe in our sample for which no early spectra have been obtained belong to this group.\\

\section{Observations}
\label{sec:observations}

The observations used here were obtained with the 48-inch Oschin Schmidt telescope at Palomar Observatory (P48), as part of the PTF project. PTF searches for transient sources by visiting selected fields of the sky regularly over the duration of several months. The data-reduction procedure is described by \cite{laher2014} and the photometric calibration by \cite{ofek2012a, ofek2012b}. We only use observations in the $R$ band and neglect the $<15$\% of data in the $g$ band to simplify this analysis. For every SN a reference period containing at least 20 observations is chosen. If possible, we use data well after the SN has faded to construct the reference image, but when no such data is available we instead resort to the oldest pre-explosion images. For PTF\,13ajn, a considerable number of observations were acquired with two CCDs, and we hence define two different reference periods. All reference periods are listed in the last column of Table \ref{tab:Samp}.

\begin{deluxetable}{llll}
\tablecolumns{4}
\tablewidth{\columnwidth}
\tablecaption{Log of spectroscopic observations}
\tablehead{
\colhead{Name}          &
\colhead{Telescope}      &
\colhead{Instrument}     &
\colhead{date}}
\startdata
PTF\,09dxv 		& Keck 1    & LRIS  & 2009-10-22 \\
SN\,2009nf 	    & Keck 1    & LRIS  & 2009-11-11 \\
PTF\,09hnq 		& HET       & LRS   & 2009-11-25 \\
PTF\,09ism 		& P200      & DBSP  & 2010-01-09 \\
PTF\,10fqg 		& P200      & DBSP  & 2010-05-07 \\
PTF\,10qrl 		& Keck 1    & LRIS  & 2010-09-05 \\
PTF\,10tzh 		& Keck 2    & DEIMOS    & 2010-10-12 \\
PTF\,10xfl 		& P200      & DBSP   & 2010-10-17 \\
SN\,2011dh 	    & Lick 3-m  & KAST   & 2011-07-06 \\
SN\,2011hg	    & Lick 3-m  & KAST   & 2011-12-19 \\
PTF\,11qju 		& Keck 1    & LRIS   & 2012-02-20 \\
SN\,2012P       & HET       & LRS    & 2012-01-14 \\
PTF\,12fxj 		& Lick 3-m  & KAST   & 2012-07-11 \\
SN\,2012ey  	& Keck 2    & DEIMOS    & 2012-10-14 \\
PTF\,12jaa 		& Keck 2    & DEIMOS    & 2012-10-14 \\
PTF\,13nu 		& P200      & DBSP      & 2013-04-13 \\
SN\,2013bb  	& Keck 1    & LRIS      & 2013-04-09 \\
PTF\,13ajn 		& P200      & DBSP      & 2013-05-02 \\
SN\,2013cu 		& P200      & DBSP      & 2013-06-02 \\
PTF\,13ebs 		& Keck 1    & LRIS      & 2013-12-02
\enddata
\tablecomments{The spectra are shown in Figures \ref{fig:spectra1}-\ref{fig:spectra3}
}
\label{tab:spec}
\end{deluxetable}

For each SN the reference image is created by coadding observations within the reference period. This image is subtracted from all science images and the (positive or negative) flux residuals at the SN location are measured using forced point-spread function (PSF) photometry. A correction for Galactic extinction according to \cite{schlegel1998} and \cite{cardelli1989} is applied to all fluxes and magnitudes in this paper.

\begin{figure*}
\centerline{\includegraphics[width=18cm]{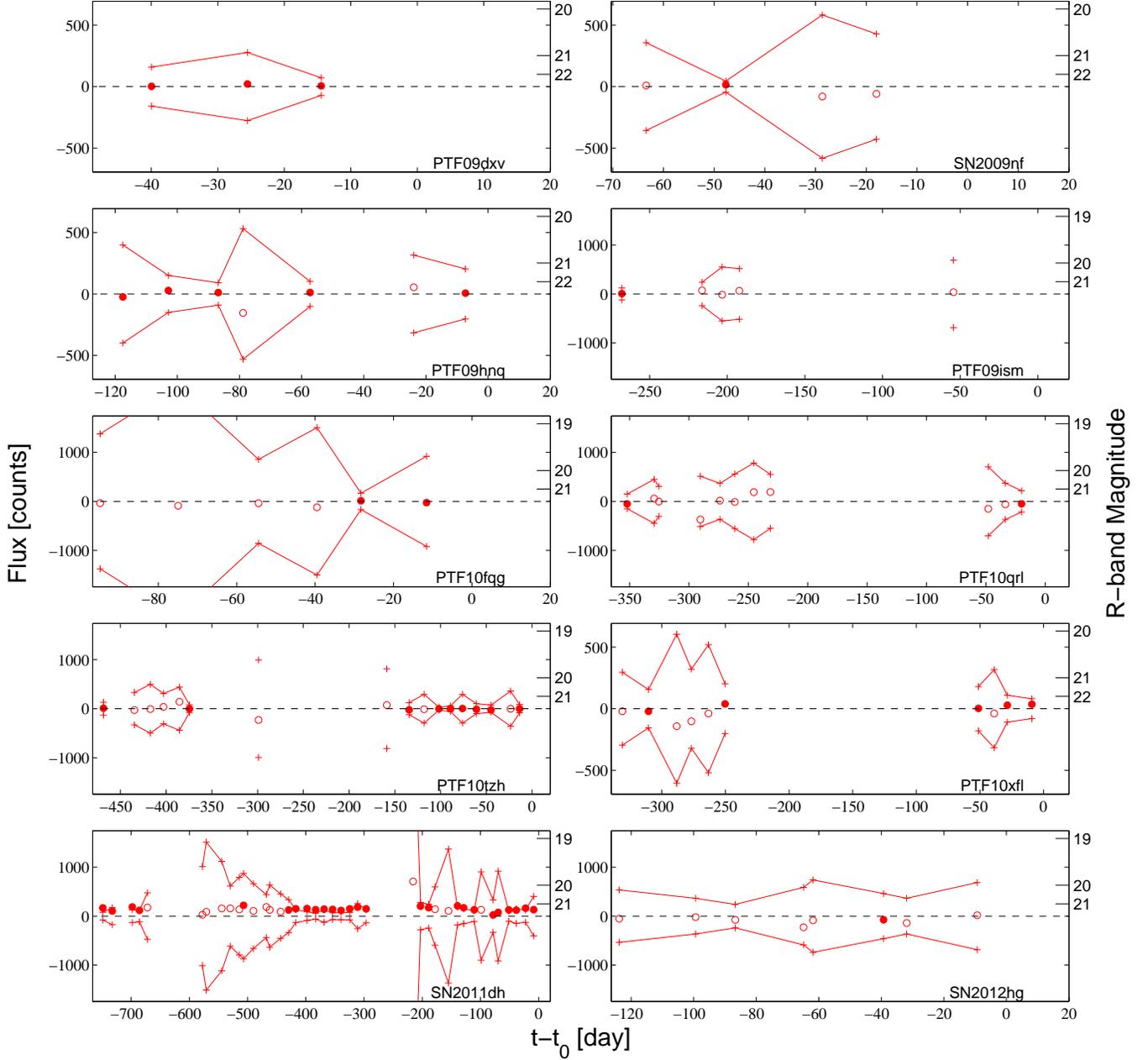}}
\caption{The pre-explosion light curves of the first ten SNe (names are listed in each panel; compare Table \ref{tab:Samp}). Shown are flux residuals (relative to the reference image given in counts per 60\,s exposure) coadded in 15-day bins prior to the SN explosion. The right-hand axis shows apparent $m_{{\rm PTF}, R}$-band magnitudes. All measurements are in the PTF magnitude system (\citealt{ofek2012a, ofek2012b}). The zero point of the flux residuals is 27. Filled circles represent 15-day bins with six or more observations and open circles mark bins having less data.
The plus signs show the 5$\sigma$ noise level, which is estimated based on scaled Poisson errors for the open circles and with the bootstrap method (\citealt{efron1982}) for filled circles (see \S\ref{sec:methods} for a detailed description). If observations are found in consecutive bins, the plus signs are connected by a solid line. A few very large error regions are outside the plotted range to improve readability. Figures \ref{fig:LCall2} and \ref{fig:LCall3} show the pre-explosion light curves of the remaining SNe.\label{fig:LCall1}}
\end{figure*}

\begin{figure*}
\centerline{\includegraphics[width=18cm]{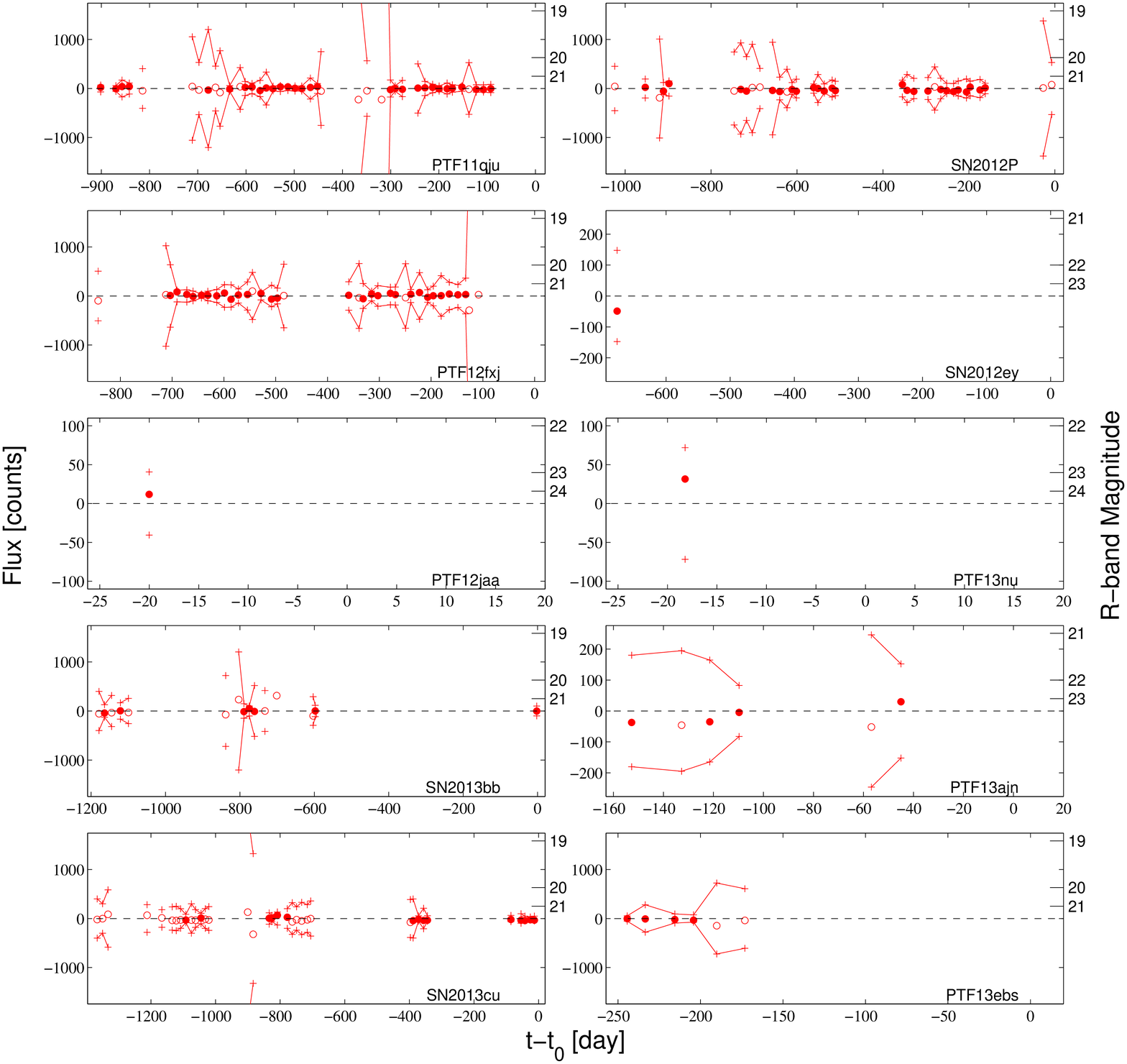}}
\caption{Figure~\ref{fig:LCall1}, continued. \label{fig:LCall2}}
\end{figure*}

\begin{figure*}
\centerline{\includegraphics[width=18cm]{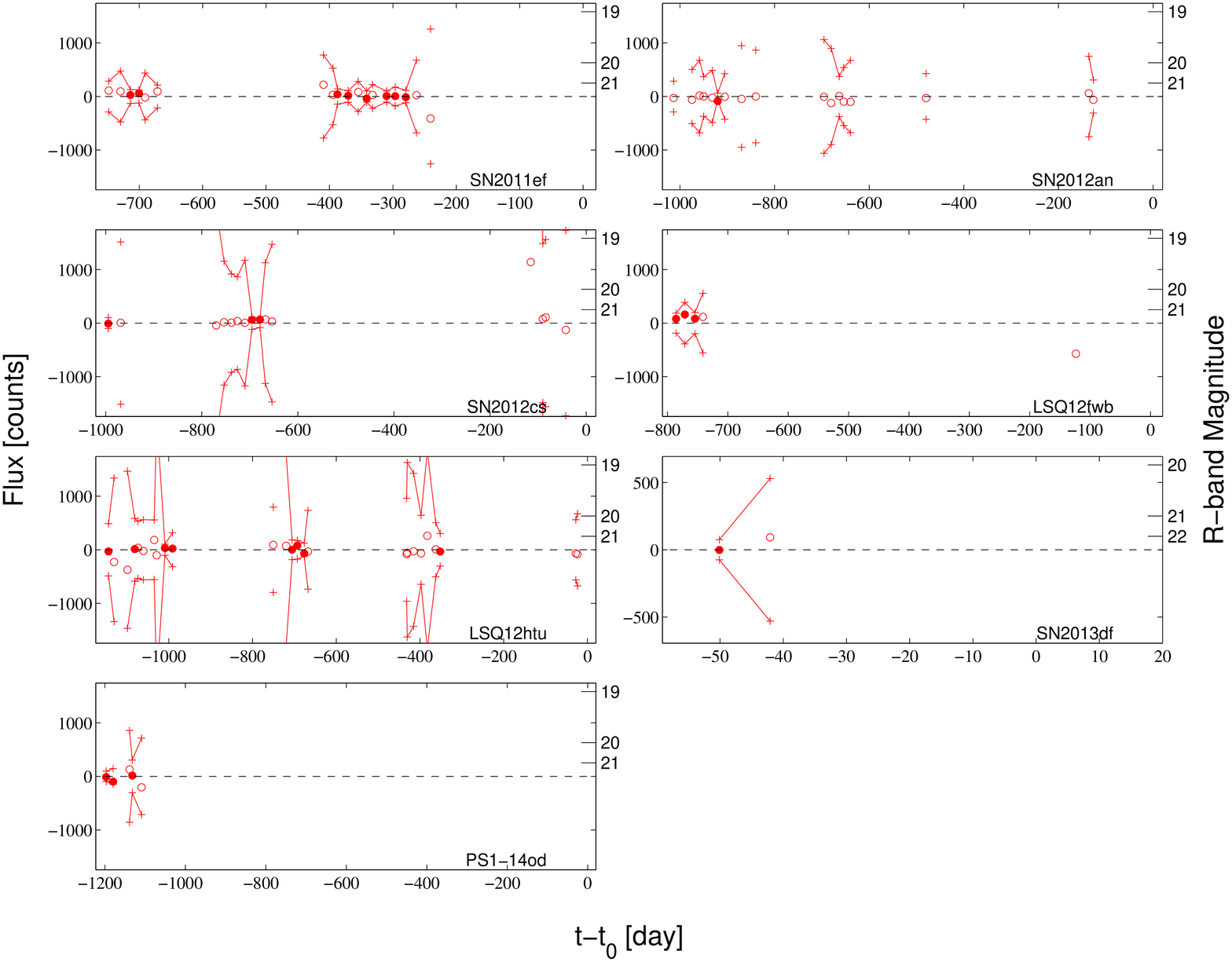}}
\caption{Figures~\ref{fig:LCall1} and \ref{fig:LCall2}, continued. Here we show the PTF pre-explosion light curves of SNe IIb found in the literature. \label{fig:LCall3}}
\end{figure*}

The observations and their flux residuals are listed in Table \ref{tab:obs}. They were obtained up to 3.5 years prior to the SN explosion. Figures \ref{fig:LCall1}--\ref{fig:LCall3} display the pre-explosion light curves in 15-day bins. Filled circles correspond to bins containing six or more observations, while open circles represent bins with fewer data points. Crosses around zero mark the estimated $5\sigma$ noise level, whose calculation is described in \S\ref{sec:methods}.\\


\section{Precursor Searches}
\label{sec:search}

The precursor search follows to a large extent the methods of \cite{ofek2014b}. In a first search we look for precursor events in individual images. In a second search channel observations are combined once in 15 and once in 60-day bins to gain sensitivity to faint precursor events. An advantage of the binned search is that it allows to estimate the noise level from the data itself which is more precise than the formal errors calculated by the image subtraction pipeline. The only drawback of binning the data is that a minimal precursor duration has to be assumed and we loose sensitivity to shorter eruptions.

Both search methods are presented in the \S\ref{sec:methods} and the results are described in \S\ref{sec:candidates}. Additional tests and cross checks on the data are explained in \S\ref{sec:tests}. This search uses tools available in the MATLAB astronomy and astrophysics package (\citealt{ofek2014a}).\\

\subsection{Search Methods}
\label{sec:methods}
In the unbinned search we have to rely on the statistical errors estimated by the image subtraction pipeline. These uncertainties are based on Poisson noise and several additional uncertainties are neglected. Those neglected error sources include errors on the fitted PSF and correlations between neighboring pixels induced by smoothing images before the subtraction with a kernel (e.g., \citealt{alard1998,bramich2008}). Therefore, the calculated errors often underestimate the fluctuations in the data and we here scale them up to obtain a more realistic error estimate.

We calculate the standard deviation of the complete pre-explosion light curve using the errors from the image subtraction pipeline and compare it to the actual scatter in the data estimated using the bootstrap method (\citealt{efron1982}). When applying the bootstrap method we resample the pre-explosion light curve 1000 times by randomnly assigning flux residuals to the observation times. Flux residuals can be drawn several times. We then calculate the average flux of every simulated light curve. The average fluxes are distributed according to a normal distribution whose standard deviation is the bootstrap error on the average flux. When using the bootstrap technique, we implicitly assume that the pre-explosion light curves are dominated by statistical fluctuations, and that eventual precursors have a negligible influence on the light curves' standard deviation. We find that the bootstrap error is always larger: For many SNe the difference is approximately a factor of two. The errors on the individual flux residuals are scaled up accordingly and we require a $5\sigma$ deviation above zero for the detection of a precursor.

We caution that these flux residuals (even in the absense of a precursor) do not follow a Gaussian distribution. Outliers can be caused in a number of ways, such as by atmospheric conditions, cosmic rays hitting the detector, or imperfect image subtractions. Large deviations are rare, however, and they can be identified by inspecting the image or verifying that a precursor candidate is detected in several subsequent images.

In the second search channel the sensitivity of the search is increased by coadding flux residuals in time bins. Our coaddition method preserves more information relative to simple image coaddition and has common grounds with the optimal coaddition method described by \cite{zackay2015}. Following \cite{ofek2014b}, we choose a bin size of 15 days, which means that precursor events having shorter durations might be missed even if their luminosity is above the quoted sensitivity.
In addition we repeat the analysis using 60-day bins to search for long-lasting faint precursors.

For the binned searches bootstrap errors are calculated for individual bins, which is more reliable than scaling up the Poisson errors estimated by the image subtraction pipeline. To get a sound error estimate with the bootstrap method, a bin must contain several entries. Therefore, in the following only bins with at least six observations are considered.

As before, we require a $5\sigma$ deviation for the detection of a precursor, with the difference that the noise level for every bin is estimated with the bootstrap method. To calculate a false-alarm probability we randomly combine observation times and flux residuals from the entire pre-explosion light curve 1000 times, bin them in 15-day bins, and look for precursors in these scrambled light curves. The probabilities for detecting one or several false precursor candidates per SN are listed in Table \ref{tab:Samp}. This test, however, only leads to a valid result if the pre-explosion light curve is dominated by statistical fluctuations and its mean flux is consistent with zero.

The binned pre-explosion light curves are shown in Figures \ref{fig:LCall1}--\ref{fig:LCall3}. In addition, bins containing fewer than six observations are marked by open circles. For these bins, the standard deviation of the mean flux is calculated based on the scaled Poisson errors.\\

\subsection{Precursor Candidates}
\label{sec:candidates}
When applying the precursor searches described in \S\ref{sec:methods} to the SNe in our sample, no precursor candidates are found in the unbinned search or when using 15-day bins, however, one precursor candidate is discovered when binning observations in 60-day bins.

In the unbinned pre-explosion light curves (not shown; flux residuals given in Table \ref{tab:obs}), no single observation has a flux above the $5\sigma$ noise level estimated based on the scaled Poisson errors. From the nondetection of events above this threshold, we can infer an upper limit on the false-alarm rate for this search channel. Using the Poisson single-sided upper limits (\citealt{gehrels1986}), the expected number of precursors is smaller than 2.3 at the 90\% confidence level when no event is observed. Since we evaluate in total 3152 pre-explosion observations, the false-alarm probability is smaller than $2.3/3152 \approx 0.0007$ per observation.

We find that even when lowering the threshold of this search to $3\sigma$, only one single flux residual reaches the threshold. The most significant subthreshold event is a $4.2\sigma$ deviation in the light curve of SN\,2012P, 654 days before the SN explosion. A second observation with similar limiting magnitude only two hours earlier, however, does not yield a detection. The two following observations 3 days later are just as deep, but neither show signs of a flare. Given the number of trials, this event is most likely spurious. We nevertheless conservatively maintain the $5\sigma$ threshold when calculating limits on the precursor rate in \S\ref{sec:rates}. Potential precursor events below the quoted flux threshold of our search do not affect the validity of the calculated limits.\\

We also do not find precursor candidates above the $5\sigma$ noise level when using 15-day bins. The only exception is the pre-explosion light curve of SN\,2011dh, the nearest object in our sample. Its progenitor star is detected in coadded images, and in 17 of the bins the noise level is low enough to allow a significant detection. The average flux of the pre-explosion light curve of SN\,2011dh is 17$\sigma$ above zero and the progenitor is hence clearly detected. In \S\ref{sec:ptf11eon} we account for the progenitor's average flux and all precursor candidates vanish, as can be seen in Figure \ref{fig:sn2011dh}.\\

\begin{figure}[bt]
\centering
\includegraphics[width=0.49\textwidth]{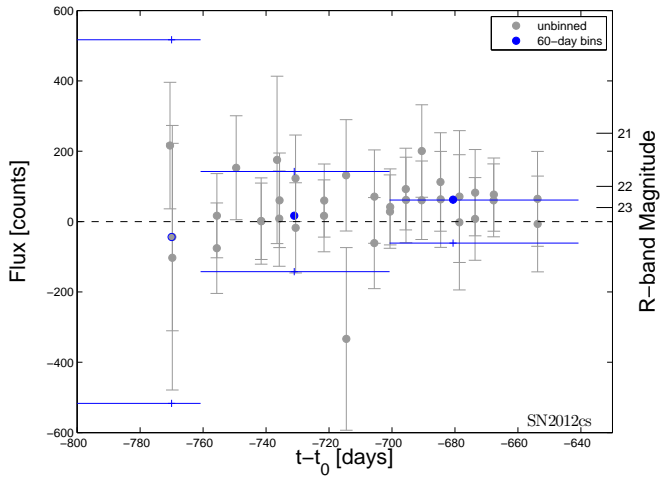}\
\caption{A section of the pre-explosion light curve of SN\,2012cs. The gray error bars show the scaled up $1\sigma$ Poisson errors on the individual observations. The blue lines indicate the bin position as well as the $5\sigma$ noise level for the binned light curves. Open dots represent bins containing fewer than 6 observations for which the noise level is estimated based on the scaled Poisson errors. Filled dots correspond to bins with at least 6 observations and the bootstrap method was used to estimate the noise level. A detection, marginally above the $5\sigma$ threshold, is obtained around day $-680$. The false alarm probability of SN\,2012cs is 1.8\% when using 60 day bins.\label{fig:sn2012cs}}
\end{figure}

When increasing the bin size to 60 days one precursor candidate is detected marginally above the $5 \sigma$ threshold in the pre-exposion light curve of SN\,2012cs. Figure \ref{fig:sn2012cs} shows the binned light curve as well as the unbinned observations. For 60-day bins the noise level is low enough to show a flux excess approximately $680$ days before the SN explosion. The 16 observations in this bin are nearly all at the same flux level. The excess is, hence, not caused by individual observations but many points contribute to it as it is expected from a long lasting precursor event. The median flux in this bin is 62 counts which corresponds to a magnitude of $m_{{\rm PTF},R}=22.51$. SN\,2012cs has a redshift of $0.0218$. The absolute magnitude of the possible precursor is $M_{{\rm PTF},R}=-11.0$ and its luminosity is $L_{{\rm prec.}}=7 \times 10^{39}$ erg\,s$^{-1}$ (no bolometric correction applied).
A duration of about 60 days seems consistent with the unbinned light curve and the radiated energy amounts to $4\times 10^{46}$ erg during this period. If real, the event is hence slightly fainter than precursors detected so far (compare e.g., \citealt{ofek2014b}) and its radiated energy is similar to the least energetic detected precursors.

Since the event is barely above the significance threshold of this search it might be a false detection. The false-alarm probability estimated for SN\,2012cs and a bin size of 60 days is 1.8\%. Its calculation is, however, based on the assumption that the pre-explosion light curve is dominated by statistical fluctuations. If the event is real, a considerable fraction of the observations (16 out of 52) are systematically shifted upwards. A real precursor can thus enhance the false-alarm probability. We conclude that the detected candidate is consistent with an astrophysical precursor event. However, due to the marginal detection we cannot decide whether or not it is real.
No further precursor candidates are found. We focus in the following on the nondetection of precursors using 15-day bins, which allows us to constrain the rate of bright precursors. The detected possible precursor is below the sensitivity threshold of this analysis and hence does not affect the limits.
The pre-explosion light curves of all 27 SNe~IIb are shown in Figures \ref{fig:LCall1}--\ref{fig:LCall3} and the upper edge of the error band corresponds to the sensitivity of the search with a bin size of 15 days.

\subsection{Additional Tests}
\label{sec:tests}

In addition to the tests described in \S\ref{sec:methods}, we check whether the pre-explosion light curves are consistent with an average flux of zero, and we look for autocorrelations in the unbinned light curves. Both tests, described below, do not reveal hints for detections or raise concerns about the data quality.

Initially, the complete pre-explosion light curves of several SNe were inconsistent with zero when calculating its average flux and its bootstrap error. In some cases the SN was still present in the chosen reference images, and we instead resorted to a pre-explosion reference. For PTF\,10qrl, we use pre-explosion images since the SN location is at the very edge of the CCD in the post-explosion images, affecting the image subtraction. Also, the average pre-explosion fluxes of several sources show deviations marginally below the $3\sigma$ threshold. This may affect the false-alarm probabilities of the binned search (shown in Table \ref{tab:Samp}). We verify, however, that no additional precursor candidates are found when subtracting the mean pre-explosion flux from all of the residuals.

Furthermore, we look for autocorrelations in the unbinned pre-explosion light curves where we calculate the correlation between an observation and the five following observations. The only deviation above the $3\sigma$ threshold is found for PTF\,10qrl with a lag of two observations. This might, however, be caused by chance. The autocorrelation is calculated for all 27 SNe with five different lags each, which means that at the $3\sigma$ level 0.36 false detections are expected. The probability to detect one or several events above the $3\sigma$ threshold is therefore $30$\%. Moreover, when looking for autocorrelations in time bins of three days instead of indexing the observations, the correlation vanishes. PTF\,10qrl is located at a redshift of 0.05 and is hence is too far away for the detection of the progenitor star. 

We conclude that no indications for precursor detections are found in these additional tests.

\section{Progenitor Detections and Variability}
\label{sec:individual}
For the three nearest SNe in our sample, here we explore the possibility of detecting the progenitor star and calculating limits on its variability.

\subsection{SN\,2011dh}
\label{sec:ptf11eon}

SN\,2011dh was discovered on 2011 May 31 in the nearby galaxy M51 at a distance of $7.8$\ Mpc, less than 15 hours after a nondetection down to a limiting magnitude of $m_{{\rm PTF}, g}=21.4$ (\citealt{arcavi2011}). The progenitor star is clearly detected both by the {\it Hubble Space Telescope} ({\it HST}; \citealt{li2011, maund2011, vandyk2011}) and in ground-based observations (e.g., the Large Binocular Telescope, \citealt{szcygiel2012}; the Northern Optical Telescope, \citealt{ergon2014a}; and PTF). The progenitor is a yellow supergiant (\citealt{vandyk2013}). Moreover, a flux excess has been measured in the fading SN light curve in blue bands and has been attributed to the presence of a binary companion star (\citealt{folatelli2014}). The companion presumably has $M_{{\rm F225W}}=-5.11\pm0.29$ mag and $M_{{\rm F336W}}=-4.66\pm0.29$ mag, consistent with a B star having $10\, {\rm M}_\odot < M < 16\, {\rm M}_\odot$ (\citealt{folatelli2014}).

There are 373 PTF $R$-band observations that span the last two years before the explosion of SN\,2011dh. Our reference is based on observations obtained starting 949 days after the explosion, when the SN has faded away. The progenitor star is hence not present in the reference image, so its flux is not subtracted from the pre-explosion light curve. All flux residuals are listed in Table \ref{tab:obs}.

\begin{figure}[t]
\centering
\includegraphics[width=0.49\textwidth]{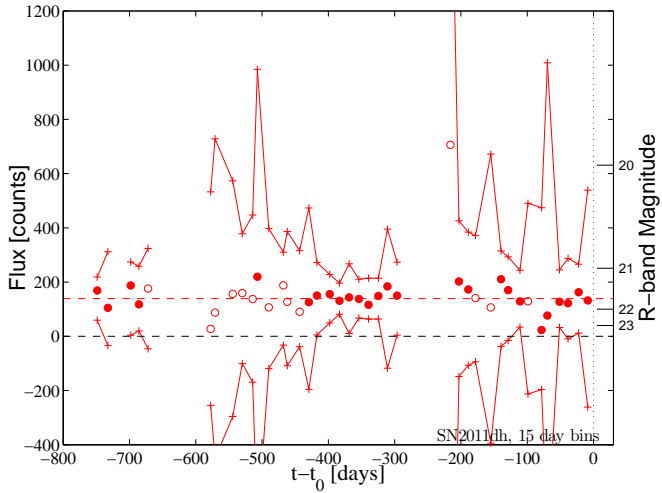}\
\caption{The binned pre-explosion light curve of SN\,2011dh. The reference images are chosen to be after the explosion, hence the flux of the progenitor has not been subtracted. No observations deviate by more than 5$\sigma$ from the mean flux.\label{fig:sn2011dh}}
\end{figure}

The progenitor is not detected in individual images; however, in the binned light curve shown in Figure \ref{fig:sn2011dh} the SN is significantly detected in many bins. The average unabsorbed flux of the progenitor is $139\pm10$ counts, corresponding to $m_{{\rm PTF}, R}=21.82\pm0.08$ mag, where the uncertainty is the $1\sigma$ bootstrap error.
During the {\it HST} observations in 2005, the progenitor star was detected in the F658N band with a magnitude of $m_{{\rm F658N}}=21.39\pm0.02$ (\citealt{vandyk2011}), corresponding to a flux of $(4.94\pm0.09)\times 10^{-18}{{\rm\ erg\ s}}^{-1}{{\rm\ cm}}^{-2}{{\rm\ \AA}}^{-1}$ at an effective wavelength of $6579.5{{\rm\ \AA}}$ (flux taken from \citealt{benvenuto2013}). This value is comparable to our measurement, translating into a flux of $(5.54\pm0.35)\times 10^{-18}{{\rm\ erg\ s}}^{-1}{{\rm\ cm}}^{-2}{{\rm\ \AA}}^{-1}$ at the same effective wavelength.

To search for precursors on top of the progenitor flux, the $5\sigma$ error band is centered around the star's average flux. No single or binned observations exceed this error band, and hence no precursors are found. One bin around day 215 before the explosion has an average flux of more than 600 counts. It however only contains two flux residuals which both have large errors. We hence conclude that this point is caused by two low-quality observations. Its deviation from zero is less than $2\sigma$.

Moreover, we search for periodic variability by calculating the power spectrum of the pre-explosion flux residuals. No significant variability was found at any time scale either in the binned or unbinned data.

\cite{szcygiel2012} analyze pre-explosion observations taken with the Large Binocular Telescope within three years before the SN explosion. With data at five different epochs they suggest a continuous fading at a rate of $0.039\pm0.006$\,mag\,yr$^{-1}$ in the $R$ band. However, in a sample of stars of comparable luminosity, as large or larger dimmings are observed for 5\% of the stars, and about 16\% of the stars have a photometric root-mean-square value larger than that of the progenitor of SN\,2011dh (\citealt{szcygiel2012}). Therefore, it is possible that this variability is not real. A fading in the last century before the SN explosion is expected as the envelope slowly responds to the rapid changes in the stellar core during the last stages of nuclear burning. The decline rates predicted by simulations are, however, typically 100 times smaller than the measured dimming rate (\citealt{szcygiel2012}).

The measured dimming of 0.039\,mag\,yr$^{-1}$ would correspond to a change of only 9 counts between our first and last observation, comparable to the uncertainty in the average flux. To check whether our observations do favor a dimming, the binned pre-explosion light curve is fitted both with a constant flux and with a dimming at the measured rate. The results are very similar, with $\chi^2/dof=35.4/39$ for the constant model compared to $\chi^2/dof=34.4/39$ for the fading light curve. With $\Delta\chi^2 \approx 1$, we cannot confirm or eliminate the fading hypothesis of \cite{szcygiel2012}.\\

\subsection{SN\,2013df}
The second-nearest object in our sample is SN\,2013df at a distance of 18.2 Mpc. No attempt can be made to detect the progenitor since we do not have potential reference images after the SN has faded. The progenitor, a $500\, {\rm R}_\odot$ cool supergiant, was identified by \cite{vandyk2014} in pre-explosion {\it HST} images with an absolute magnitude of $M_{{\rm HST}, V} \approx -6.8$.\\

\subsection{SN\,2012P}
\label{sec:ptf12os}
At the position of SN\,2012P, a potential progenitor star was detected in archival {\it HST} observations from 2005 March 10 (\citealt{vandyk2012}). The source has an absolute magnitude of $M_{{\rm HST}, V} \approx -9.1$, and its luminosity and colors are consistent with a very luminous supergiant star or a star cluster.

Interpolating the $V$ and $I$-band magnitudes reported by \cite{vandyk2012}, a PTF $R$-band magnitude of $m_{{\rm PTF}, R}=22.74$ is expected, corresponding to a signal of 50 counts per observation. The weighted average flux of the 308 pre-explosion observations is, however, $-17\pm10$ counts per observation ($1\sigma$ bootstrap error). Hence, we conclude that the source reported by \cite{vandyk2012} is still present in the reference image, which consists of observations more than 779 days after the SN explosion. The source close to the SN position could be either a compact star cluster as mentioned by \cite{vandyk2012}, a luminous binary companion, or an unrelated object.

Based on the PTF pre-explosion observations in the last three years before the explosion of SN\,2012P, we can exclude precursor explosions with an absolute $R$-band magnitude of $M_{{\rm PTF}, R}\leq -11$ ($L\ge 8\times 10^{39} {{\rm erg\ s}}^{-1}$ without a bolometric correction) during 30\% of the time.\\

\section{Control Time and Precursor Rates}
\label{sec:rates}

Following \cite{ofek2014b}, we combine the results of the individual SNe and calculate an upper limit on the precursor rate of SNe~IIb. Hereby, we implicitly assume that the progenitors of SNe~IIb form a uniform population and that our SN sample provides a good representation of this population. Both presumptions are likely not fulfilled and the resulting rate estimate should hence be regarded with caution.

To address the inhomogeneity of the  progenitor population, we additionally calculate an upper limit only for SNe~IIb with observed double-peaked light curves since this is a strong indication for an extended envelope (\citealt{nakar2014}).

Moreover, faint SNe~IIb are likely underrepresented in this sample; however, our sample does cover a wide range of peak luminosities and light-curve shapes. This analysis is thus not restricted to any certain subgroup of SNe~IIb.
The advantage of a rate estimate is that we potentially gain information on the SNe~IIb as a class and we are able to quantify the results of this analysis.

To calculate a precursor rate for all SNe~IIb, we estimate the control time --- the time during which a precursor above a certain absolute magnitude can be detected. The following calculations are described for a bin size of 15 days. When using 60-day bins all calculations and results are similar except for the detection of the faint precursor candidate. All resulting upper limits or precursor rates are summarized in Table \ref{tab:rates} for both bin sizes.

We adopt a minimal precursor duration of 15 days and bin the observations accordingly. For bins with six or more observations, the 1$\sigma$ noise level is calculated using the bootstrap method and is then multiplied by a factor of 5 to estimate a $5\sigma$ level. For bins with fewer observations, we determine the standard deviation based on the scaled Poisson errors estimated by the image-subtraction pipeline (see \S\ref{sec:methods}). The upper edge of the noise level corresponds to the $5\sigma$ limiting magnitude per bin. The control time is formally given as
\begin{equation}
\begin{split}
t_{\rm search}(M_{R})&= \sum^{{N_{\rm bin}},N_{{\rm SN}}}_{i=1} t_{i{\rm , search}} (M_{i,R}) \\
{\rm with\ }\quad t_{i,{\rm search}}(M_{i,R})&=
 \begin{cases}
 15 {{\rm\ days,}}& {{\rm if\ }} M_{i,R}<M_{R}\\
 0 {{\rm\ days,}}& {{\rm otherwise,}}\\
 \end{cases}
 \end{split}
\end{equation}
where $M_{R}$ is the absolute magnitude at which the control time is calculated and $M_{i,R}$ is the $5\sigma$ limiting magnitude in bin $i$. The sum runs over all $N_{\rm bin}$ light-curve bins per SN and all $N_{\rm SN}$ SNe in this sample. The mean observation time of the bin relative to the explosion date, the limiting magnitude, and the number of observations per bin are listed in Table \ref{tab:controltime} (unbinned observations given in Table \ref{tab:obs}).

\begin{figure}[htb]
\includegraphics[width=0.49\textwidth]{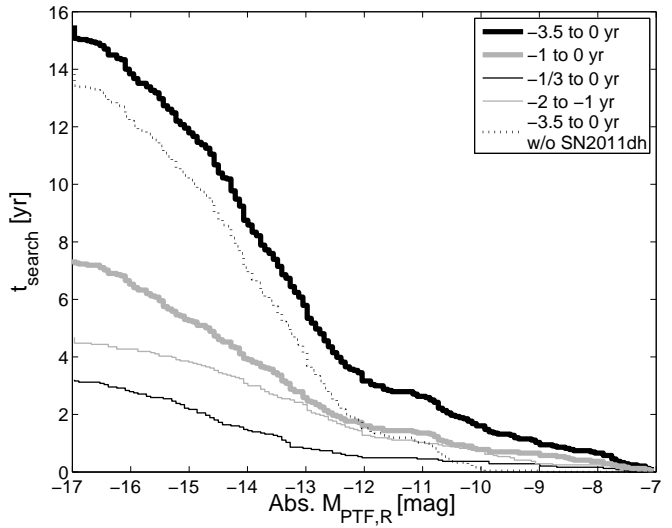}
\caption{\label{fig:controltime}The absolute-magnitude dependent control time for the complete sample, during which the SN locations were observed and no precursors were detected above the given magnitude. The various curves are for precursors taking place at different time ranges prior to the SN explosion (see legend). The curve normalization and shape are determined by the amount of data and the SN distances. SN\,2011dh accounts for most observations at the brightest magnitudes and the dotted line shows the effect of removing it from the sample. Additional curves display the control time for shorter periods before the explosion. The limiting magnitudes per bin for all SNe are listed in Table \ref{tab:controltime}.}
\end{figure}

From the search results described in \S\ref{sec:candidates} and the light curves shown in Figures \ref{fig:LCall1}--\ref{fig:LCall3}, we know that during the control time no precursors were detected above the estimated noise level when using 15-day bins. The control time $t_{\rm search}$ for the complete sample as a function of the precursor magnitude is depicted in Figure \ref{fig:controltime}, where $t_{\rm search}$ is defined as the bin duration of 15 days multiplied by the number of bins in which a precursor of the respective magnitude would have been detected. For example, at an absolute magnitude of $-14$ and considering the full time range of 3.5 years prior to the SN explosion (thick black line in Figure \ref{fig:controltime}), a control time of $9.2$ years has been accumulated; in our sample there are 223 15-day bins in which precursors having an absolute magnitude of $-14$ or less can be excluded.

According to the amount of data and the SN distance, the objects in our sample contribute differently to the magnitude-dependent control time. The five SNe with the longest control time (PTF\,11qju, SN\,2011dh, PTF\,12fxj, SN\,2012P, and SN\,2011ef) account for close to 50\% of the total time covered by observations. Especially at large absolute magnitudes, the control-time distribution is dominated by the two nearby events, SN\,2011dh and SN\,2012P. In Figure \ref{fig:controltime} we show how the distribution changes when removing SN\,2011dh, which owing to its small distance has a large contribution to the control time in the absolute magnitude range $-7$ to $-10$. In a similar way, the second bump at magnitudes from $-9$ to $-12$ is due to the observations of SN\,2012P. At lower magnitudes many other SNe start to contribute to the distribution, so the results are less dominated by individual objects.

\begin{deluxetable}{lllll}
\tablecolumns{5}
\tablewidth{\columnwidth}
\tablecaption{Precursor Search Control Time}
\tablehead{
\colhead{Name} &
\colhead{ $\Delta t$ } &
\colhead{ $m_{{\rm PTF}, R}$ } &
\colhead{Abs. $M_{{\rm PTF}, R}$} &
\colhead{$N_{{\rm meas}}$}\\
\colhead{ } &
\colhead{(day)} & 
\colhead{(mag)} & 
\colhead{(mag)} & 
\colhead{ }
}
\startdata
       PTF\,09dxv &   $-40.0$& 21.49& $-14.28$ &  20\\
       PTF\,09dxv &   $-25.5$& 20.90& $-14.87$ &  23\\
       PTF\,09dxv &   $-14.4$& 22.38& $-13.39$ &  14\\
       SN\,2009nf &   $-63.4$& 20.61& $-15.96$ &   4\\
       SN\,2009nf &   $-47.7$& 22.87& $-13.70$ &   8\\
       SN\,2009nf &   $-28.6$& 20.08& $-16.49$ &   3\\
       SN\,2009nf &   $-18.0$& 20.41& $-16.16$ &   3\\
       PTF\,09hnq &  $-117.5$& 20.50& $-14.88$ &   9\\
       ...
\enddata
\tablecomments{\label{tab:controltime}Listed are the SN name, the mean observation time of the 15-day bin relative to the explosion date, the apparent and absolute magnitude down to which precursors can be excluded, and the number of observations per bin. The limiting magnitudes are at the 5$\sigma$ level estimated from Poisson errors for bins with less than 6 observations and with the bootstrap method otherwise (\S\ref{sec:methods}). They correspond to the upper $+$ signs in Figures \ref{fig:LCall1}--\ref{fig:LCall3}, and their cumulative distribution is shown in Figure \ref{fig:controltime}. The full table is available in the online version.}
\end{deluxetable}

\begin{center}
\begin{figure*}[tb]
\subfigure[90\% upper limit on precursor rate derived from the complete SN~IIb  sample\label{fig:limit}]{\includegraphics[width=0.49\textwidth]{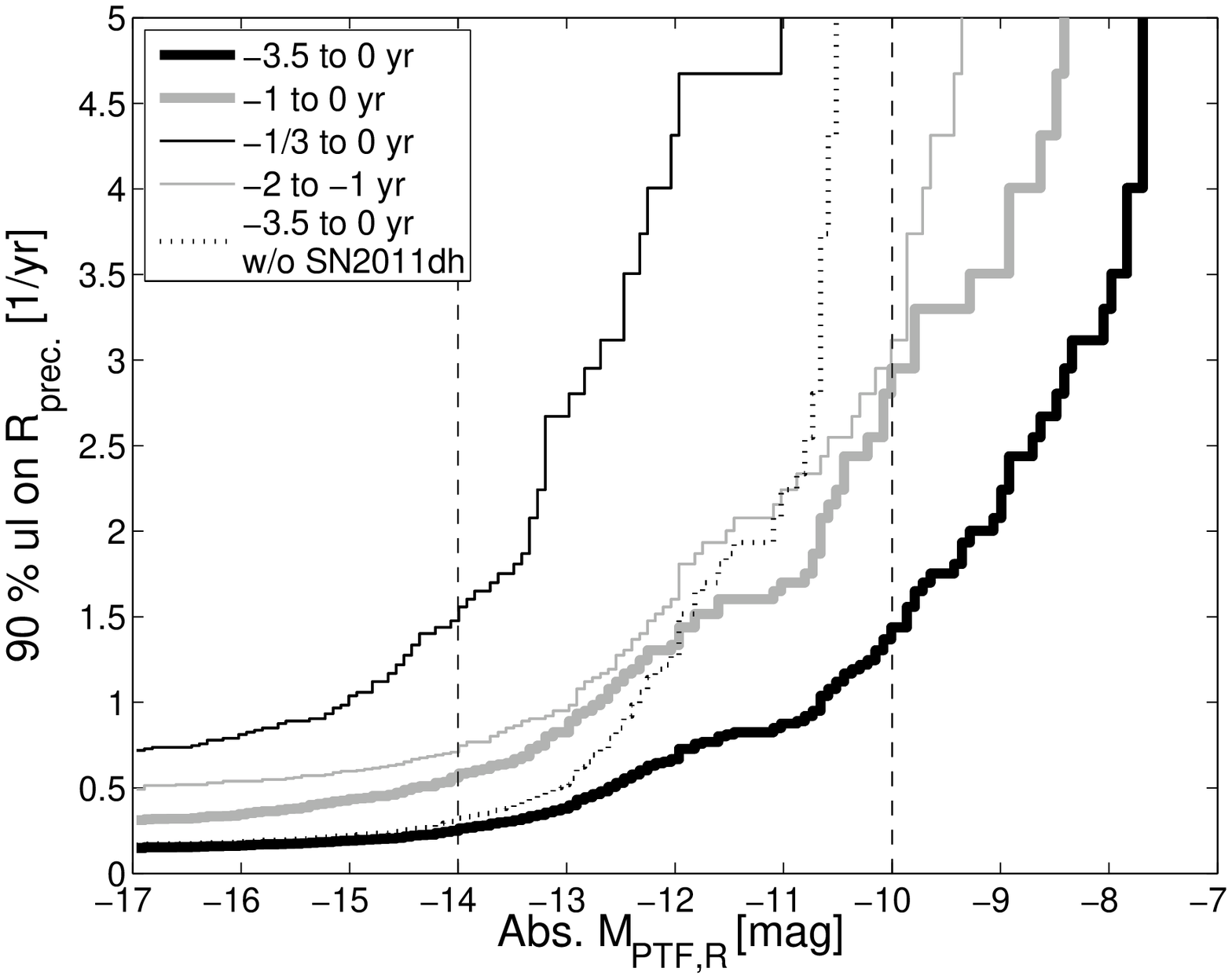}}
\hfill
 \subfigure[90\% upper limit on precursor rate derived from five double-peaked SNe~IIb\label{fig:limit_double}]{\includegraphics[width=0.49\textwidth]{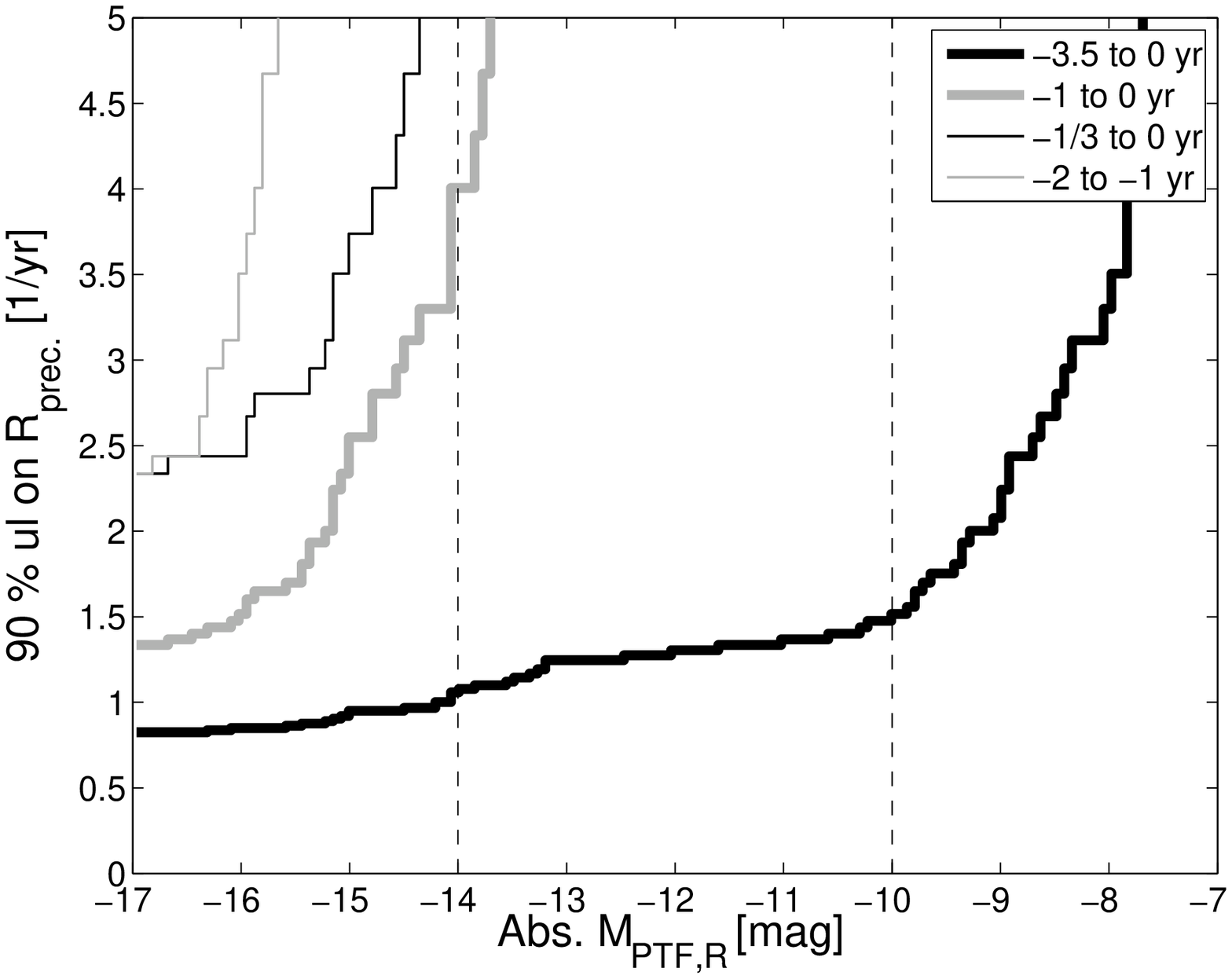}}
\caption{The upper limits on the average precursor rate per SN calculated based on the control time shown in Figure \ref{fig:controltime}. At the lowest (brightest) magnitudes the limit is dominated by observations of SN\,2011dh. Considering only double-peaked SNe, the limits are weaker owing to the smaller amount of data.}
\end{figure*}
\end{center}

\begin{deluxetable*}{llllll}
\tablecolumns{6}
\tablewidth{\textwidth}
\tablecaption{$N_{\rm prec}$, average number of precursors per SN}
\tablehead{
%\colhead{Time interval} & \multicolumn{4}{c}{$N_{\rm prec}$, average number of precursors per SN} \\
\colhead{Time interval} & \multicolumn{2}{c}{SNe IIb, bright precursors ($-14$ mag)} & \multicolumn{2}{c}{SNe IIb, faint precursors ($-10$ mag)} & \colhead{SNe IIn ($-14$ mag)}\\
\colhead{(yr)} & \colhead{15-day bins} & \colhead{60-day bins} & \colhead{15-day bins}& \colhead{60-day bins} & \colhead{15-day bins}}
\startdata
$-3.5$ to $0$ 	& $<0.86$ 	& $<0.51$	& $<5.03$	& $1.52^{+3.50, +7.12}_{-1.26, -1.49}$	& - \\
$-2.5$ to $0$ 	& - 		& - 		& -		& -					& $4.8^{+3.8, +7.6}_{-2.3, -3.5}$ \\
$-1$ to $0$ 	& $<0.56$ 	& $<0.36$	& $<2.95$	& $<2.00$				& $2.8^{+2.2, +4.4}_{-1.3, -2.1}$\\
$-1/3$ to $0$ 	& $<0.52$ 	& $<0.29$	& $<2.08$	& $<2.33$				& $2.5^{+2.0, +4.0}_{-1.2, -1.8}$\\
$-2$ to $-1$ 	& $<0.70$ 	& $<0.47$	& $<3.12$	& $1.01^{+2.33, +4.75}_{-0.84, -0.99}$	& - 
%$-3.5$ to $0$ 	& $<$0.86 	& $<$5.03	& - & $<0.51$ & $0.23^{+0.52, +1.06}_{-0.17, -0.22}$ \\
%$-2.5$ to $0$ 	& - 		& - 		& $4.8^{+3.8, +7.6}_{-2.3, -3.5}$ & - & - \\
%$-1$ to $0$ 	& $<$0.56 	& $<$2.95	& $2.8^{+2.2, +4.4}_{-1.3, -2.1}$ & $<0.36$ & $<2.00$\\
%$-1/3$ to $0$ 	& $<$0.52 	& $<$2.08	& $2.5^{+2.0, +4.0}_{-1.2, -1.8}$ & $<0.29$ & $<2.33$\\
%$-2$ to $-1$ 	& $<$0.70 	& $<$3.12	& - & $<0.47$ & $0.22^{+0.50, +1.02}_{-0.18, -0.21}$ 
\enddata
\tablecomments{\label{tab:rates} Upper limit on allowed number of precursors $N_{\rm prec}$ within the respective time interval (rates from Figure \ref{fig:limit} multiplied by time interval) compared to measured number precursors for SNe~IIn (\citealt{ofek2014b}). The limits apply to precursors with a minimal duration of 15 and 60 days respectively. The constraints on long precursors are stronger since more observations are coadded which leads to a better sensitivity. All limits are at the $90$\% confidence level and the quoted uncertainties correspond to the $1\sigma$ and $2\sigma$ Poisson errors (\citealt{gehrels1986}).}
\end{deluxetable*}
For searches in which no precursors are detected, we can set a 90\% upper limit on the expectation number of precursors of 2.3 (\citealt{gehrels1986}; one-sided Poisson statistics). The limits on the average precursor rate per SN per year, $R_{\rm prec}$, and on the average number of precursors per SN, $N_{\rm prec}$, are given by 
\begin{eqnarray}
 R_{\rm prec} < \frac{2.3}{t_{\rm search}} \qquad {{\rm and\ }} \qquad N_{\rm prec} < 2.3 \frac{t_{\rm window}}{t_{\rm search}},
\end{eqnarray}
respectively, where $t_{\rm window}$ is the duration of the considered time interval before the explosion. The limit on $N_{\rm prec}$ is hence purely determined by the ratio $t_{\rm window}/t_{\rm search}$, the amount of time during which precursors can be excluded compared to the time window during which the search is conducted. We note that $R_{\rm prec}$ is an average quantity since the bins of $t_{\rm search}$ are not necessarily distributed homogenously within $t_{\rm window}$.

The resulting upper limit on the precursor rate is shown in Figure \ref{fig:limit}, and Table \ref{tab:rates} compares the upper limit on the number of precursors per SN to the number of precursors measured for SNe~IIn by \cite{ofek2014b}.

The upper limit on the number of precursors for the longest time interval of 3.5 years prior to the SN explosion is smaller than unity. This means that at the 90\% confidence level, not all SNe~IIb can exhibit precursors as bright as $-14$ mag, which is a typical $R$-band luminosity of precursors detected so far (e.g., \citealt{pastorello2007,ofek2014b}). Compared to SNe~IIn (\citealt{ofek2014b}), the average number of precursors prior to SNe~IIb at magnitude $-14$ is lower by at least a factor of five (see Table \ref{tab:rates}). In case precursors are equally common before SNe~IIb and SNe~IIn, they have to be at least as dim as an absolute magnitude of $-10$ for SNe~IIb, about a factor of 40 dimmer compared to the precursors observed prior to SNe~IIn (see Table \ref{tab:rates}).

As explained in the introduction, a double-peaked light curve presumably requires the presence of an extended hydrogen envelope (\citealt{bersten2012,nakar2014}). The duration of the first peak is usually only hours to few days and most SNe are not discovered as early. For most SNe in our sample we thus do not know whether or not an early peak is present. However, not all SNe~IIb have an early peak (\citealt{arcavi2015}), and there are indications that some SN~IIb progenitors might be compact (\citealt{chevalier2010}). To account for this diversity in the progenitor population, we additionally calculate limits considering only the data of the five SNe~IIb for which two peaks are observed by PTF or reported in the literature. As stated in Table \ref{tab:Samp}, this includes PTF\,10tzh, SN\,2011dh, PTF\,12jaa, PTF\,13ajn, and SN\,2013df (see also \citealt{arcavi2015}). These SNe contribute only 15\% of the control time of the complete sample, and the limits are hence drastically weakened, as shown in Figure \ref{fig:limit_double}.\\

\section{Discussion}
\label{sec:discussion}

In this section we estimate whether the explosive ejection of a low-mass stellar envelope in a precursor event is likely to be bright enought to be detectable in this analysis. We expect a radiatively efficient precursor eruption to radiate an amount of energy comparable to (or larger than) the binding energy of the ejected envelope, which is given by

\begin{equation}\label{eq:binding_energy}
 E_{{\rm bind}} \approx G\ \frac{M_{{\rm env}}\ M_{{\rm core}}}{R_{{\rm core}}},
\end{equation}
where $G$ is the gravitational constant, $M_{\rm env}$ and $M_{\rm core}$ are respectively the masses of the envelope and the core, and $R_{\rm core}$ is the core radius above which the envelope is located prior to its ejection.

According to Nakar \& Piro (2014; see also \citealt{piro2015}), both $M_{\rm env}$ and $R_{\rm core}$ can be estimated from the shape of the first peak in the optical light curve. $M_{\rm env}$ is determined by the time at which the bolometric light curve reaches the first peak ($t_p$), and $R_{\rm core}$ can be derived from the minimal luminosity $L_{\rm min}$ between the two peaks. $L_{\rm min}$ can be obtained relatively precisely whenever a first peak is observed. The rise time $t_p$ is usually less well constrained, since it can be as short as a day or several hours, and in most cases the data are is not sufficient.

In our sample, the only double-peaked SN for which a good upper limit on $t_p$ is available is SN\,2011dh, where the maximum of the first peak was reached at most 15 hours after the explosion (\citealt{arcavi2011}). This limit refers to the peak in visible light. However, the bolometric peak probably takes place even earlier, hence this limit holds.
For SN\,2011dh, numerical simulations by \cite{bersten2012} suggest $M_{\rm core}\approx 4\, {\rm M}_\odot$, $M_{\rm env}\approx 0.003\, {\rm M}_\odot$, and $R_{\rm core}\approx 5\times10^{11} \, {\rm cm}$, comparable to the values estimated by \cite{nakar2014}.
Using Eq. \ref{eq:binding_energy}, the binding energy is  $E_{\rm bind} \approx 8 \times 10^{45} {\rm\ erg}$. This result is an order-of-magnitude estimate since most quantities entering the calculation have large uncertainties of up to a factor of a few. The kinetic energy required to unbind the envelope relates to the bolometric energy released in the precursor event, $E_{\rm prec}$, via an unknown radiation efficiency factor $\epsilon$: $E_{\rm rad, prec}=\epsilon \ E_{\rm bind}$ (see \citealt{ofek2015}).

In the case of SN\,2011dh, within 2 years prior to the SN explosion, 50\% of the time is covered by observations with a limiting magnitude of $-9$ or less and gaps no longer than two weeks. This limits the precursor luminosity during this time to $L_{\rm prec} < 10^{39}\,{\rm erg\,s^{-1}}$ (no bolometric correction applied). If the envelope of this progenitor star is unbound during our observations (50\% probability), we can set the following limit on the radiative efficiency $\epsilon$, as a function of the precursor duration $\delta t$, where the emitted luminosity is assumed to be constant over the precursor duration:
\begin{equation}
 \epsilon = \frac{E_{{\rm rad, prec}}}{E_{{\rm bind}}} \approx \frac{L_{{\rm prec}} \ \delta t}{E_{{\rm bind}}} \lesssim 0.16 \left( \frac{\delta t}{15 \,{{\rm\ days}}} \right).
\end{equation}
The unbinding event (if it ever existed) might be too faint to be detected if the efficiency $\epsilon$ is low or if the precursor lasts for several months or longer. Another possibility is that the envelope is ejected prior to the time interval probed in this analysis.

The limit on the precursor luminosity $L_{\rm prec}$ is close to the Eddington luminosity for a $4\,{\rm M}_\odot$ progenitor. Thus, a continuum-driven wind (e.g.,\citealt{shaviv2001}), which could in principle explain giant eruptions of luminous blue variable stars or SN~IIn precursors, is likely ruled out as an explanation for the Type IIb SN envelope ejection. Stellar winds can, however, be driven by other mechanisms such as line absorption or absorption by dust which do not require as high luminosities; see \citet{langer2012} for a review.

If the precursor candidate detected prior to SN\,2012cs is real it released an energy of $\sim4\times10^{46}$ erg (see \S\ref{sec:candidates}). We do not have much information about the progenitor star. However, if it is similar to the progenitor of SN\,2011dh the radiated energy is similar to the required energy for the unbinding event. It is hence possible that a stellar envelope was ejected in this event.

\section{Summary}
\label{sec:summary}
We examine the pre-explosion light curves of 27 nearby SNe~IIb, searching for outbursts. One precursor candidate is marginally detected in a single 60-day bin in the pre-explosion light curve of the nearby SN\,2012cs. The probability to measure such an event caused by noise is 1.8\% for this pre-explosion light curve. The possible precursor happened 680 days before the SN explosion and if real its absolute $R$-band magnitude is $M_{R{\rm ,PTF}}=-11.0$, which corresponds to a luminosity of $7\times10^{39}$ erg\,s$^{-1}$ (no bolometric correction applied) and a radiative energy release of $4\times10^{46}$ erg within the approximate duration of 60 days.

When binning the observations in 15-day bins no precursor eruptions are found, and we calculate a magnitude-dependent limit on the average precursor rate among SNe~IIb. Precursors as luminous as $-14$ magnitude occur on average $<0.86$ times within the final 3.5 years before the SN explosion, while in the last year the average number of precursors is limited to $<0.56$ at 90\% confidence level. These limits are obtained under the assumption that precursor eruptions last for at least 15 days. We conclude that bright precursor explosions, if they occur at all, are rare and do not happen before the explosion of every SN of Type IIb.

By contrast, precursors are frequent for SNe~IIn (\citealt{ofek2014b}); at the $99$\% confidence level, $98$\% of all SNe~IIn have precursor eruptions brighter than absolute magnitude $-14$ within the $2.5$ years before the explosion.
The precursor rate of SNe~IIn measured by \cite{ofek2014b} exceeds the upper limit for SNe~IIb by about a factor of five at an absolute magnitude of $-14$ or less. In addition, the precursor rate for SNe~IIb can be constrained at higher magnitudes of up to $-7$.

For the nearby SN~IIb SN\,2012P, our observations show that the source detected in pre-explosion {\it HST} observations at the SN position (\citealt{vandyk2012}) is still present more than two years after the SN explosion and thus cannot be the progenitor star.

The progenitor of SN\,2011dh, the closest SN in our sample, is clearly detected in coadded observations, and its mean $R$-band magnitude is $m_{{\rm PTF}, R}=21.82\pm0.08$ mag, consistent with archival {\it HST} observations. The possible slow fading reported by \cite{szcygiel2012} is below the sensitivity threshold of our observations. However, with 373 observations over the last two years prior to the SN explosion the progenitor is monitored nearly constantly, and no variability or precursors are detected. We argue that for this progenitor star, the ejection of the stellar envelope in a precursor event might be observable except if the process is radiatively inefficient or lasts over several months.

\acknowledgments

This paper is based on observations obtained with the Samuel Oschin
Telescope as part of the Palomar Transient Factory project, a
scientific collaboration between the California Institute of
Technology, Columbia University, Las Cumbres Observatory, the Lawrence
Berkeley National Laboratory, the National Energy Research Scientific
Computing Center, the University of Oxford, and the Weizmann Institute
of Science. We are grateful for excellent staff assistance at
Palomar and Lick Observatories. E.O.O. is incumbent of the Arye
Dissentshik career development chair and is grateful to support by
grants from the Willner Family Leadership Institute Ilan Gluzman
(Secaucus NJ), Israeli Ministry of Science, Israel Science Foundation,
Minerva and the I-CORE Program of the Planning and Budgeting Committee
and The Israel Science Foundation.  A.G.-Y. is supported by the EU/FP7
via ERC grant no. 307260, the Quantum Universe I-Core program by the
Israeli Committee for planning and budgeting, and the ISF, Minerva and
ISF grants, WIS-UK ``making connections'' and the Kimmel and ARCHES
awards. M.S. acknowledges support from the Royal Society and EU/FP7-
ERC grant no [615929]. N.J.S. thanks the IBM Einstein Fellowship
support by the Institute for Advanced Study, Princeton.  A.V.F's
research was made possible by National Science Foundation grant
AST-1211916, the TABASGO Foundation, and the Christopher R. Redlich
Fund.

%\clearpage

\end{document}